\documentclass[12pt]{article}
\usepackage[english]{babel}
\usepackage[utf8]{inputenc}
\usepackage[T1]{fontenc}

\usepackage{amsmath}

\usepackage{amsfonts}
\usepackage{url}
\usepackage[dvips]{graphicx}
\usepackage{calc}
\usepackage{subfigure}
\usepackage{multirow}

\def\mL{\mathcal{L}}
\def\bD{\mathbf{D}}
\def\mH{\mathcal{H}}

\def\bx{\mathbf{x}}
\newcommand{\pb}[1]{\left\{#1\right\}}

\def\tX{\tilde{X}}

\def\mG{\mathcal{G}}
\def\by{\mathbf{y}}
\def\bz{\mathbf{z}}

\def\bT{\mathbf{T}}

\def\mM{\mathcal{M}}

\begin{document}
	\begin{titlepage}
	\begin{center}
		{\Large{ \bf Canonical Analysis of Gravity with Dynamical Determinant of Metric-General Case}}
		
		\vspace{1em}  
		
		\vspace{1em} J. Kluso\v{n} 			
		\footnote{Email addresses:
			klu@physics.muni.cz (J.
			Kluso\v{n}) }\\
		\vspace{1em}
		\textit{Department of Theoretical Physics and
			Astrophysics, Faculty of Science,\\
			Masaryk University, Kotl\'a\v{r}sk\'a 2, 611 37, Brno, Czech Republic}
		
		\vskip 0.8cm
		
		%
		%
		%
		%
		%
		%
		
		\vskip 0.8cm
		
	\end{center}

\begin{abstract}
We analyse general form of theory with the dynamical determinant of metric. We show that due to the presence of general function of determinant that multiplies scalar
curvature 	Hamiltonian constraint is either second class constraint or it is necessary to impose  condition of transversality on parameters of spatial diffeomorphism. 
\end{abstract}

\bigskip

\end{titlepage}

\newpage

\section{Introduction and Summary}\label{first}
Theories with restricted diffeomorphism invariance attached great interest recently as an alternative solution of the cosmological constant problem. Famous example of  such theories are unimodular theories of gravity where the determinant of the metric is fixed \cite{Buchmuller:1988wx,Henneaux:1989zc,Kuchar:1991xd,Unruh:1988in}. Clearly if we demand that the action is invariant under change of coordinates we find that it is not invariant under full diffeomorphism but instead under restricted one which preserves determinant of the metric. 
The restriction imposed on allowed diffeomorphism transformations has non-trivial consequences
on the canonical structure of theory as was shown for example in 
\cite{Karataeva:2022mll,Bufalo:2017tms,Bufalo:2015wda,Kluson:2014esa}.

More precisely, General Relativity is invariant under general diffeomorphism (Diff) transformations. On the other hand it was shown long time ago in 
\cite{vanderBij:1981ym}
that symmetry group for a consistent
description of the massless graviton can be maximal subgroup of Diff that is known as 
TDiff since the parameter that characterizes the infinitesimal form of 
diffeomorphism transformation is transverse one which means that it obeys the condition
\begin{equation}
\partial_\mu \xi^\mu=0 \ . 
\end{equation}
As we argued above characteristic property of theories invariant under restricted diffeomorphism 
is the fact that determinant of metric transforms as scalar under diffeomorphism transformation 
rather than as scalar density. Then there is a natural question whether it is possible to consider theories where determinant of metric appears as ordinary scalar. 
Such a model was firstly discussed in 
\cite{Alvarez:2006uu} and the most general form of such theory was presented in 
\cite{Lopez-Villarejo:2010uib}. The quantum behaviour of such  a theory was studied in 
\cite{Lopez-Villarejo:2010uib} and confrontation of this theory with observation was 
analyzed in \cite{Alvarez:2009ga}. 

The fact that the most general form of TDiff invariant theory contains kinetic term for determinant
makes it also very interesting from the Hamiltonian point of view. In our previous paper 
\cite{Kluson:2023bkg} we studied special form of this action where the function in front of the scalar curvature is equal to square root of the determinant of metric. We found corresponding Hamiltonian and identified primary and secondary constraints. We also showed that Hamiltonian constraint is tertiary constraint which arises from the requirement of the preservation of 
secondary constraints. Then we calculated Poisson brackets between these constraints and we showed that they closed. In other words the Hamiltonian constraint is the first class constraint too. 

In this paper we would like to study the most general form of TDiff invariant action when we consider general function of determinant of metric in front of the scalar curvature in the action.
It turns out that the presence of this general function has crucial impact on the consistency of theory. In more details, in order to proceed to the canonical formalism we should implement 
$3+1$ decomposition of scalar curvature, for review see 
\cite{Gourgoulhon:2007ue}. Then presence of terms proportional to covariant derivatives in this expansion has crucial impact on the canonical structure of theory as we show in the bulk of this paper. More precisely, we find Hamiltonian that contains kinetic terms for spatial components of metric and for lapse while there are only $n-1$ primary constraints (in $n-$dimensional space-time) corresponding to the vanishing conjugate momenta to shift functions. Then the requirement of  preservation of the primary constraints leads to an emergence of $n-1$ secondary constraints which are spatial diffeomorphism constraints. We calculate Poisson brackets between these constraints
and canonical variables. Finally the requirement of the preservation of these constraints leads to the Hamiltonian constraint. In other words, the Hamiltonian for TDiff invariant gravity is again given as  sum of constraints. The crucial difference is in the fact that when we calculate Poisson bracket between Hamiltonian constraints we find that it is not equal to linear combinations of spatial diffeomorphism constraints.  Rather we find that it is equal to some complicated functions of canonical momenta. This result have two interpretations. The first one is based on the presumption that time evolution of the system is governed by the original Hamiltonian with primary constraints included only. In this case the requirement of the preservation of Hamiltonian constraint would implies an existence of additional constraint. Then we should again study time evolution of this quaternary constraint and so on. We should then expect an infinite number of constraints to be generated which is clearly non-consistent. 

The second interpretation is based on an existence of extended Hamiltonian which is defined as Hamiltonian where all constraints are included 
\cite{Henneaux:1992ig}. In this case the fact that Poisson bracket between smeared form of Hamiltonian constraints does not vanish on constraint surface implies that Hamiltonian constraint is second class constraint with themselves. This is not so surprising as it sounds if we recognize
that Hamiltonian constraint corresponds to $\infty^{n-1}$ constraints defined in $\mathbf{R}^{n-1}$. Moreover, similar situation occurs in case of non-projectable
Ho\v{r}ava-Lifshitz gravity \cite{Horava:2009uw}, where it was shown in case when lapse is space-dependent that the theory is inconsistent \cite{
Blas:2009yd,Li:2009bg,Henneaux:2009zb} in the sense that the Hamiltonian constraints are second class constraints among themselves. The situation was solved by including
appropriate kinetic term for lapse \cite{Blas:2009qj}, see also \cite{Kluson:2010nf},
however it is not clear whether similar procedure can be applied for the model studied in this paper. 

As we wrote above requirement of the preservation of spatial diffeomorphism constraints led to an emergence of Hamiltonian constraint however this is not single possibility. We show that the requirement of the preservation of spatial diffeomorphism constraint can be solved by imposing restriction on the parameters of spatial diffeomorphism, namely $\partial_i\xi^i=0$ where $i=1,\dots,n$. Note that this condition can be imposed since $\mH_T$ is scalar not scalar density which is sharp difference with respect to general relativity case. Further, the spatial diffeomorphism transverse condition implies that there is no Hamiltonian constraint and hence all problems that were mentioned in previous paragraph are eliminated. On the other hand if we combine TDiff condition $\partial_\mu\xi^\mu=0$ with $\partial_i\xi^i=0$ we find that $\xi^0$ does not depend on time. In some way this can be interpreted as satisfactory result since there is no Hamiltonian constraint on the other hand it suggests that Hamiltonian formalism cannot reproduce TDiff constraint $\partial_\mu\xi^\mu=0$. 

Let us outline our results. We found Hamiltonian for general TDiff invariant theory which is characterized by two general functions $F(g),G(g)$ where no restrictions on the form of these functions were imposed. We identified primary and secondary constraints and we argued that the requirement of the preservation of secondary constraint leads either to the spatial diffeomorphism transverse condition or to an emergence of tertiary Hamiltonian constraint. We argued that the second possibility leads to the theory with not well defined canonical structure. On the other hand imposing the first possibility leads to consistent theory but with absent time reparametrization. We mean that both these problems deserve further investigation 
and hence theories invariant under TDiff can serve as good laboratories for the application of canonical formalism on more complicated theories than the fully diffeomorphism invariant ones. 

This paper is organized as follows. In the next section (\ref{second}) we introduce theory with dynamical determinant of metric and find corresponding Hamiltonian. 
Then in section (\ref{third}) we study stability of constraints and discuss conditions when the theory is well defined.

\section{General TDiff-Invariant Action}
\label{second}
In this section we introduce basic formulation of gravity with time dependent metric. 
This action has a form \cite{Lopez-Villarejo:2010uib}
\begin{eqnarray}\label{Sgen}
&&S=\int d^{n}x\mL \ , \quad 
\mL=\frac{1}{\kappa} F(\sqrt{-g})\sqrt{-g}[R(g)+\frac{G(\sqrt{-g})}{\sqrt{-g}g^2}\partial_\mu g^{\mu\nu}
\partial_\nu g] \ , \quad \nonumber \\
&& g\equiv \det g \ , \quad \kappa=16\pi \ , 
\end{eqnarray}
where $F$ and $G$ are general functions of $\sqrt{-g}$. Note that we work in 
$n$-dimensional space-time with metric signature $(-,+,\dots,+)$ and $\mu,\nu=0,1,\dots,n$. 
 Since under general 
transformations $x'^\mu=x^\mu+\xi^\mu(x)$ the metric tensor transforms as 
\begin{equation}
g'_{\mu\nu}(x')=g_{\mu\nu}(x)-g_{\mu\sigma}(x)\partial_\nu\xi^\sigma(x)-
\partial_\mu \xi^\sigma g_{\sigma\nu}(x) 
\end{equation}
it is clear that the determinant transforms as 
\begin{equation}
g'(x')=g(x)-\partial_\mu\xi^\mu g(x) \ .
\end{equation}
Then clearly the action (\ref{Sgen}) is not invariant under full diffeomorphism but instead under restricted ones where parameters $\xi^\mu$ obey the conditions
\begin{equation}
\partial_\mu \xi^\mu=0 \ . 
\end{equation}
Due to the presence of kinetic term for $g$ it is clearly very interesting
to find Hamiltonian form of the action (\ref{Sgen}). Note that the special case of the action (\ref{Sgen}) with $F(\sqrt{-g})=1$ was studied recently in 
\cite{Kluson:2023bkg}. In this paper we would like to relax this condition and consider  general function $F$. We will see that it will lead to important consequences.

To proceed to the canonical formulation we use the well known $3+1$
formalism that is the fundamental ingredient of the Hamiltonian
formalism of any theory of gravity \footnote{For recent review, see
	\cite{Gourgoulhon:2007ue}.}. We consider $n$ dimensional manifold
$\mathcal{M}$ with the coordinates $x^\mu \ , \mu=0,\dots,n-1$ and
where $x^\mu=(t,\bx) \ , \bx=(x^1,x^2,\dots,x^{n-1})$. We presume that this
space-time is endowed with the metric $g_{\mu\nu}(x^\rho)$
with signature $(-,+,\dots,+)$. Suppose that $ \mathcal{M}$ can be
foliated by a family of space-like surfaces $\Sigma_t$ defined by
$t=x^0=\mathrm{const}$. Let $h_{ij}, i,j=1,2,\dots,n-1$ denotes the metric on $\Sigma_t$
with inverse $h^{ij}$ so that $h_{ij}h^{jk}= \delta_i^k$. We further
introduce the operator $\nabla_i$ that is covariant derivative
defined with the metric $h_{ij}$.
We also define  the lapse
function $N=1/\sqrt{-g^{00}}$ and the shift function
$N^i=-g^{0i}/g^{00}$. In terms of these variables we
write the components of the metric $g_{\mu\nu}$ as
\begin{eqnarray}
g_{00}=-N^2+N_i h^{ij}N_j \ , \quad g_{0i}=N_i \ , \quad
g_{ij}=h_{ij} \ ,
\nonumber \\
g^{00}=-\frac{1}{N^2} \ , \quad g^{0i}=\frac{N^i}{N^2} \
, \quad g^{ij}=h^{ij}-\frac{N^i N^j}{N^2} \ .
\nonumber \\
\end{eqnarray}
and hence $g=-N^2\det h_{ij}$. We further have following decomposition of $R$ in the form
\begin{eqnarray}
&&R=K^{ij}K_{ij}-K^2+r+2\tilde{\nabla}_\mu[\tilde{n}^\mu K]
-\frac{2}{N}\nabla_i\nabla^iN \ , 
\nonumber \\
&&K_{ij}=\frac{1}{2N}(\partial_t h_{ij}-\nabla_i N_j-\nabla_j N_i) \ , 
\quad n^0=\sqrt{-g^{00}} \ , \quad 
n^i=-\frac{g^{0i}}{\sqrt{-g^{00}}} \ , \nonumber \\
\end{eqnarray}
and where $r$ is scalar curvature defined with $h_{ij}$ and where $\tilde{\nabla}_\mu$ is covariant derivative compatible with $g_{\mu\nu}$ so that $\tilde{\nabla}_\mu g_{\rho\sigma}=0$ while 
$\nabla_i$ is covariant derivative compatible with the metric $h_{ij}$. Note that we can also write
\begin{eqnarray}
&&	\tilde{\nabla}_\mu [n^\mu K]=\frac{1}{\sqrt{-g}}\partial_\mu[\sqrt{-g}n^\mu K]
\ , \nonumber \\
&&\frac{1}{g^2}\partial_\mu g g^{\mu\nu}\partial_\nu g=
-\frac{1}{g^2}\frac{1}{N^2}(\partial_0 g-N^i\partial_i g)^2+
\frac{1}{g^2}h^{ij}\partial_i g\partial_j g \ . \nonumber \\
\end{eqnarray}
As in our previous paper
\cite{Kluson:2023bkg}
 we perform following manipulation 
with the kinetic term for determinant of metric
\begin{eqnarray}
\frac{1}{Ng}(\partial_0 g-N^i\partial_i g)
=\frac{2}{N^2}\partial_0 N+2\partial_i (\frac{N^i}{N})+2K_{ij}h^{ji} \ 
\nonumber \\
\end{eqnarray}
and also 
\begin{equation}
\partial_0 h_{ij}h^{ij}=2N K_{ij}h^{ji}+2\nabla_i N_jh^{ji} \ . 
\end{equation}
Then the action (\ref{Sgen})
has the form  
\begin{eqnarray}\label{act}
&&S=\frac{1}{\kappa}\int d^nx [F\sqrt{-g}K_{ij}\mG^{ijkl}K_{kl}
+\sqrt{-g}Fr+2F\partial_\mu (N\sqrt{h}n^\mu K)-2F\sqrt{h}\nabla_i\nabla^iN-\nonumber \\
&&-\frac{G}{g^2N^2}(\partial_0 g-N^i\partial_i g)^2+\frac{G}{g^2}
h^{ij}\partial_i g\partial_j g]=\nonumber \\
&&=\frac{1}{\kappa}\int d^nx \left[F\sqrt{-g}K_{ij}\mG^{ijkl}K_{kl}
+F\sqrt{-g}r
+\frac{FG}{g^2}
h^{ij}\partial_i g\partial_j g
+\right.\nonumber \\
&&-4F'g\sqrt{-g}\left(\frac{1}{N^2}\partial_0N+\partial_i(\frac{N^i}{N})+K\right) K-\nonumber \\
&&\left.-4FG\left(\frac{1}{N^2}\partial_0N+2\partial_i(\frac{N^i}{N})+K\right)^2
-2F\nabla_i\nabla^i\sqrt{-g}\right] \ , 
\nonumber \\
\end{eqnarray}
where $\mG^{ijkl}$ is defined as
\begin{equation}
\mG^{ijkl}=\frac{1}{2}(h^{ik}h^{jl}+h^{il}h^{jk})-h^{ij}h^{kl} \ . 
\end{equation}
The crucial point of the canonical analysis is that the  action (\ref{act})  can be rewritten into following form 
\begin{eqnarray}\label{StX}
S=\frac{1}{\kappa}\int d^nx 
[\sqrt{-g}FK_{ij}\mM^{ijkl}K_{kl}-4FG\tX^2+F\sqrt{-g}r+\frac{FG}{g^2}h^{ij}\partial_i g\partial_j g
-2F\nabla_i \nabla^i\sqrt{-g}] \ ,  \nonumber \\ 
\end{eqnarray}
where we defined $\tX$ as 
\begin{equation}
\tX=\frac{\partial_0 N}{N^2}+\partial_i(\frac{N^i}{N})+\left(1+\frac{1}{2}\frac{F'g\sqrt{-g}}{FG}\right)K  \  \quad 
\end{equation}	
and metric $\mM^{ijkl}$ in the form 
\begin{equation}
\mM^{ijkl}=\mG^{ijkl}-\frac{F'^2g^3}{F^2G\sqrt{-g}}h^{ij}h^{kl}=
\frac{1}{2}(h^{ik}h^{jl}+h^{il}h^{jk})-(1+\frac{F'^2g^3}{F^2G\sqrt{-g}})
h^{ij}h^{kl} 
 \ . 
\end{equation}
From the action (\ref{StX})  we get conjugate momenta
\begin{eqnarray}\label{momenta}
&&\pi^{ij}=\frac{\sqrt{h}F}{\kappa }\mM^{ijkl}K_{kl}-\frac{8}{\kappa}FG\tX(1+\frac{1}{2}\frac{F'g\sqrt{-g}}{FG})
\frac{1}{2N}h^{ij} \ , \nonumber \\
&&\pi_N=\frac{\partial \mL}{\partial \partial_0N}=-\frac{8}{\kappa}FG\frac{1}{N^2}
\tX \ . \nonumber \\	
\end{eqnarray}
Using these relations we obtain Hamiltonian density in the form
\begin{eqnarray}\label{mHK}
&&\mH=\pi^{ij}\partial_0 h_{ij}+\pi_N\partial_0 N-\mL=\nonumber \\
&&=\frac{1}{\kappa}\sqrt{-g}FK_{ij}\mM^{ijkl}K_{kl}-
\frac{4}{\kappa}FG\tX^2-\sqrt{-g}Fr-\nonumber \\
&&-\frac{FG}{g^2}h^{ij}\partial_i g\partial_j g
+2F\nabla_i\nabla^i\sqrt{-g} +N^i\mH_i \ , \nonumber \\
\end{eqnarray}
where we implicitly used integration by parts to introduce $\mH_i$ as
\begin{equation}
\mH_i=-2\nabla_j \pi^{jk}h_{ki}+\partial_i\pi_NN+
2\pi_N\partial_iN \ . 
\end{equation}
It is clear that we have to express Hamiltonian density in terms of canonical
variables $\pi^{ij},\pi_N$ instead of $K_{ij}$ and $\partial_0N$. To do this we use 
relation between $\tX$ and $\pi_N$ given in (\ref{momenta}) to replace $\tX$ in the definition of $\pi^{ij}$ to get
\begin{eqnarray}\label{defPi1}
\pi^{ij}-\frac{1}{2}(1+\frac{1}{2}\frac{F'g\sqrt{-g}}{FG})\pi_NNh^{ij}=
\frac{\sqrt{h}}{\kappa }F\mM^{ijkl}K_{kl} \ . \nonumber \\ 
\end{eqnarray}
In this paper we will presume that the matrix $\mM^{ijkl}$ is non-singular with inverse matrix $\mM_{ijkl}$ that obeys the relation 
\begin{equation}\label{defMinverse}
\mM_{ijkl}\mM^{klmn}=\frac{1}{2}(\delta_i^m\delta_j^n+\delta_i^n\delta_j^m)  \ . 
\end{equation}
With this presumption we can express $K_{ij}$ from (\ref{defPi1}) as 
\begin{eqnarray}\label{Kinv}
K_{ij}=\frac{\kappa}{\sqrt{h}F}\mM_{ijkl}\Pi^{kl} \ , \quad 
\Pi^{ij}=\pi^{ij}-\frac{1}{2}(1+\frac{1}{2}\frac{F'g\sqrt{-g}}{FG})\pi_NN h^{ij}\ . 
\nonumber \\
\end{eqnarray}
Before we continue further we determine explicit form of the matrix $\mM_{ijkl}$. Let us presume that it has the form
\begin{equation}
\mM_{ijkl}=\frac{1}{2}(h_{ik}h_{jl}+h_{il}h_{jk})-\bD h_{ij}h_{kl} \ . 
\end{equation}
Then from (\ref{defMinverse}) we can determine $\bD$ to be equal to
\begin{eqnarray}
\bD=\frac{(1+\frac{F'^2g^3}{F^2G\sqrt{-g}})}{(n-1)
(1+\frac{F'^2g^3}{F^2G\sqrt{-g}})-1} \ . 
\nonumber \\
\end{eqnarray}
Note that for $F'=0$, $\bD=-\frac{1}{n-2}$ and hence $\mM_{ijkl}=\mG_{ijkl}
=\frac{1}{2}(h_{ik}h_{jl}+h_{il}h_{jk})-\frac{1}{n-2}h_{ij}h_{kl}$ that is well known form of de Witt inverse metric. 
Then  using (\ref{Kinv}) in (\ref{mHK}) we obtain final form of Hamiltonian 
 as function of canonical variables 
\begin{eqnarray}
&& H=\int d^{n-1}\bx \mH \ , \nonumber \\
&& \mH=\frac{\kappa}{\sqrt{h}F} N\Pi^{ij}\mM_{ijkl}\Pi^{kl}-\frac{\kappa}{16}\frac{1}{FG}N^4\pi_N^2
-N\sqrt{h}Fr-\frac{FG}{g^2}h^{ij}\partial_i g\partial_j g+\nonumber \\
&&+2F\nabla_i \nabla^i\sqrt{-g} +N^i\mH_i
\equiv \mH_T+N^i\mH_i \ .
\nonumber \\	
\end{eqnarray}
This is final form of Hamiltonian for TDiff invariant gravity. 
\section{Stability of Constraints}\label{third}
Since original Lagrangian does not contain time derivative of shift functions $N^i$ it is clear that corresponding conjugate momenta are absent. Alternatively we say that there are primary constraints of theory
\begin{equation}\label{pbN}
\pi_i(\bx)\approx 0 \ , \quad \pb{N^i(\bx),\pi_j(\by)}=\delta^i_j\delta(\bx-\by) \ . 
\end{equation} 
By definition the constraint is stable if it is preserved during time evolution of the system
\begin{equation}
\partial_t \pi_i(\bx)=\pb{\pi_i(\bx),H}=
-\mH_i(\bx) \approx 0
\end{equation}
using Poisson brackets given in (\ref{pbN}).  We see that the requirement of the preservation of constraint $\pi_i\approx 0$ implies secondary constraints 
\begin{equation}
\mH_i(\bx)\approx 0 \ . 
\end{equation}
For further purposes we introduce its smeared form defined as
\begin{equation}
\bT_S(\xi)=\int d^{n-1}\bx\xi^i\mH_i
\end{equation}
and calculate  Poisson brackets between $\bT_S(\xi)$ and fundamental fields
\begin{eqnarray}\label{PBfun}
&&\pb{\bT_S(\xi),h_{ij}}=-\partial_m h_{ij}\xi^m-\partial_i \xi^m
h_{mj}-h_{im}\partial_j \xi^m \ , \nonumber \\
&&\pb{\bT_S(\xi),N}=-\xi^i\partial_i N+\partial_i \xi^i N \ , \nonumber \\
&&\pb{\bT_S(\xi),\pi^{ij}}
=-\partial_m(\xi^m \pi^{ij})+\partial_m\xi^i\pi^{mj}+
\pi^{im}\partial_m\xi^j \ , \nonumber \\
&&\pb{\bT_S(\xi),\pi_N}=	-\partial_i\pi_N\xi^i-2\pi_N\partial_i \xi^i \ , 
\nonumber \\
&&\pb{\bT_S(\xi) h}=-\partial_mh\xi^m-2\partial_i\xi^i h \ , \nonumber \\
&&
\pb{\bT_S(\xi),N^2h}=-
\xi^m\partial_m (N^2h) \ , \nonumber \\
&&\pb{\bT_S(\xi),r}=-\xi^m\partial_m r \ . \nonumber \\
\end{eqnarray}
Using these basic Poisson brackets we also get
\begin{eqnarray}\label{PBfun2}
&&\pb{\bT_S(\xi^i), g}=
\pb{\bT_S(\xi^i),N\sqrt{g}}=-\xi^m\partial_m g \ , \nonumber \\
&&\pb{\bT_S(\xi^i),\Pi^{ij}}
=-\partial_m\xi^m\Pi^{ij}-\xi^m\partial_m\Pi^{ij}+\partial_m\xi^i\Pi^{mj}+\Pi^{im}\partial_m\xi^j \ . \nonumber \\
\end{eqnarray}
The first line in (\ref{PBfun2})  gives an important result that says that $g$ behaves as scalar under 
spatial diffeomorphism transformations. Further, the expression on the second line
in (\ref{PBfun}) suggests  that
$N$ behaves as tensor density of weight $-1$. This fact is supported by following
observation.  Since $\sqrt{-g}=N h$ behaves as scalar under spatial diffeomorphism 
it is natural to demand that 
\begin{equation}
	\nabla_i(\sqrt{-g})=\nabla_i (N\sqrt{h})=\partial_i (\sqrt{h}N) \ . 
\end{equation}
On the other hand from definition of covariant derivative ($\nabla_i h=0$) we get
\begin{eqnarray}
	\nabla_i N=\frac{1}{\sqrt{h}}(\partial_i(N\sqrt{h}))
	=\frac{1}{\sqrt{h}}(\partial_i N\sqrt{h}+N\partial_i\sqrt{h})=
	\partial_iN+\Gamma^k_{ki}N  \nonumber \\
\end{eqnarray}
which is consistent with the Poisson bracket between $\bT_S(\xi)$ and $N$. 
Then we can write 
\begin{eqnarray}
	F\nabla_i\nabla^i\sqrt{-g}=F\nabla_i[h^{ij}\partial_j\sqrt{-g}]=
	F\frac{1}{\sqrt{h}}\partial_i[\sqrt{h}h^{ij}\partial_j\sqrt{-g}]
	\nonumber \\
\end{eqnarray}
and hence
\begin{eqnarray}
	\pb{\bT_S(\xi),F\nabla_i \nabla^i\sqrt{-g}}
=-\xi^m\partial_m \left[\frac{F}{\sqrt{h}}\partial_i[\sqrt{h}h^{ij}\partial_j \sqrt{-g}]\right] \ . 
\nonumber \\
\end{eqnarray}
Now collecting all terms together we get that 
\begin{eqnarray}
\pb{\bT_S(\xi),\mH_T}=-\xi^m\partial_m \mH_T \ 
\nonumber \\
\end{eqnarray}
that shows that $\mH_T$ behaves as scalar under spatial diffeomorphism transformations.  Note that this is an important difference
with respect to the Hamiltonian constraint in General relativity which is 
scalar density. 
Now we can study time evolution of the constraint $\mH_i\approx 0$ or its smeared
form. Note that the  time evolution of  constraint $\bT_S(\xi)$ has the form 
\begin{eqnarray}
&&\partial_t \bT_S(\xi)=\pb{\bT_S(\xi),H_T}=\int d\by\pb{\bT_S(\xi),\mH(\by)}+
\pb{\bT_S(\xi),\bT_S(N^j)}=\nonumber \\
&&=-\int d\by\xi^m\partial_m\mH_T
+\bT_S((\xi^j\partial_j N^i-N^j\partial_j\xi^i)) \nonumber \\
\end{eqnarray}
using the fact that 
\begin{eqnarray}
\pb{\bT_S(\xi^i),\bT_S(N^j)}=\int d^{n-1}\bx (\xi^j\partial_j N^i-N^j\partial_j\xi^i)\mH_i
=\bT_S((\xi^j\partial_j N^i-N^j\partial_j\xi^i)) \ . \nonumber \\
\end{eqnarray}
We see that the constraint $\mH_i\approx 0$ is preserved on condition when following 
expression vanishes 
\footnote{It is instructive to compare this result with the situation when $\mH_T$ is scalar density when we get $\int d^{n-1}\bx (-\xi^m\partial_m\mH_T-\mH_T\partial_m\xi^m)=-\int d^{n-1}\bx\partial_m(\xi^m\mH_T)$ which vanishes with appropriate boundary conditions. In other words requirement of the preservation of constraint $\bT_S(\xi)$ does not generate new constraint in case
when $\mH_T$ is scalar density.}
\begin{equation}
\int d^{n-1}\bx \xi^m\partial_m\mH_T=0 \ .  
\end{equation}
There are two possibilities how to make this expression zero. The first one is to perform integration by parts and then we obtain condition on the diffeomorphism parameters in the form 
\begin{equation}
\partial_m\xi^m=0 \ . 
\end{equation}
The second possibility is to impose tertiary constraint
\begin{equation}
\mH_T\approx 0 \ . 
\end{equation}
We will discuss each condition individually. 

\subsection{Condition $\partial_m \xi^m=0$}
Imposing the condition $\partial_m \xi^m=0$ implies that there are no 
constraint equivalent to the Hamiltonian one. This fact implies that there is an additional physical mode with respect to General relativity which  is $N$ 
and  conjugate momentum. However there is a tension with the Lagrangian formalism where we showed that the action is invariant under restricted space-time diffeomorphism $\partial_\mu \xi^\mu=0$. In fact, if we combine this condition with 
$\partial_m\xi^m=0$ we obtain that
\begin{equation}
\partial_0 \xi^0=0 \  
\end{equation}
this implies that $\xi^0=\xi^0(\bx)$. On the other hand time reparametrization is gauge symmetry in case when $\xi^0$ depends on time as well. In other word the condition that $\xi^0$ does not depend on $t$ is consistent with the absence of the Hamiltonian constraint. 

\subsection{Imposing  Constraint $\mH_T\approx 0$}
Let us consider the  second possibility which is introducing another constraint $\mH_T\approx 0$. Then we should clearly check  whether this constraint is preserved during the time evolution of the system. To do this we introduce smeared form of this constraint
\begin{equation}\label{smearHam}
\bT_T(X)=\int d\bx X\sqrt{h}\mH_T \ , \quad \bT_T(Y)=\int d\by  Y\sqrt{h} \mH_T \ , 
\end{equation}
where  $X$ and $Y$ are ordinary test functions and where we  include factor $\sqrt{h}$ into definition of these constraints since $\mH_T$ is scalar.  In principle it is straightforward to calculate Poisson bracket between these constraints (\ref{smearHam}) however the result is very complicated. In order to demonstrate it let us calculate  following expression 
\begin{eqnarray}\label{Pimmr}
&&\pb{\int d^{n-1}\bx X\sqrt{h}\frac{\kappa}{\sqrt{h}F}N\Pi^{ij}\mM_{ijkl}\Pi^{kl}, 
	-\frac{1}{\kappa}\int d^{n-1}\by \sqrt{h}N\sqrt{h}Fr}+\nonumber \\
&&+\pb{-\frac{1}{\kappa}\int d^{n-1}\bx X\sqrt{h} N\sqrt{h}Fr,\int d^{n-1}\by Y \sqrt{h}\frac{\kappa}{\sqrt{h}F}
	\Pi^{ij}\mM_{ijkl}\Pi^{kl}}=
\nonumber \\
&&=-4\int d^{n-1}\bx (Y\nabla_mX-X\nabla_mY)N^2h\Pi^{mn}\frac{\nabla_nF}{F}+\nonumber \\
&&+4\int d^{n-1}\bx (Y\nabla_m X-X\nabla_mY)N^2hh^{mn}\frac{\nabla_nF}{F}\Pi(1-\bD(n-2))+
\nonumber \\
&&+2\int d^{n-1}\bx (Y\nabla_mX-X\nabla_mY)N^2h\nabla_n\Pi^{mn}-
\nonumber \\
&&-2\int d^{n-1}\bx (Y\nabla_mX-X\nabla_mY)h^{mn}N^2h\nabla_n(\Pi(1-\bD(n-2)))
\nonumber \\
\end{eqnarray}	
using this formula
\begin{equation}
	\frac{\delta r(\bx)}{\delta h_{ij}(\by)}=-r^{ij}(\bx)\delta(\bx-\by)
	+\nabla^i\nabla^j\delta(\bx-\by)-h^{ij}(\bx)\nabla_k\nabla^k\delta(\bx-\by) \ 
\end{equation}
and  Poisson brackets
\begin{eqnarray}
&&\pb{\Pi^{ij}(\bx),h_{kl}(\by)}=-\frac{1}{2}(\delta^i_k\delta^j_l+\delta^i_l
\delta^j_k)\delta(\bx-\by) \ , \nonumber \\
&&\pb{\Pi^{ij}(\bx),g(\by)}=\frac{1}{2}\frac{F'g^2\sqrt{-g}}{FG}h^{ij}\delta(\bx-\by) \ .
\nonumber \\
\end{eqnarray}
The result given in (\ref{Pimmr}) demonstrates the crucial point of the calculations of Poisson brackets
between smeared forms of $\mH_T$. We see that there are additional  terms with respect to the
similar calculations performed in case of  general relativity. 
The first one is proportional to derivative of $F$. The next two ones are proportional to $1-\bD(n-2)$ which are generally space-time dependent. Note that this 
expression is zero in case of General Relativity when $\bD=\frac{1}{n-2}$. 
In the similar way we should proceed with the calculations of remaining Poisson brackets and we find that these terms do not cancel each other. Instead we get following  schematic result
\begin{equation}
\pb{\bT_T(X),\bT_T(Y)}=\int d^{n-1}\bx(X\partial_m Y-Y\partial_mX)\mathbf{M}^m \ , 
\end{equation}
where $\mathbf{M}^m$ is complicated expression that does not vanish on the constraint surface.
This  result  has following important consequence for the time evolution of the constraint $\mH_T$
\begin{eqnarray}
&&\partial_t\mH_T(\bz)=\pb{\mH_T(\bz),H}\approx 
\int d^{n-1}\by \pb{\mH_T(\bz),\mH_T(\by)}=
\nonumber \\
&&
=\frac{2}{\sqrt{h}}\partial_m[\frac{1}{\sqrt{h}}]\mathbf{M}^m+
\frac{1}{h}\partial_m\mathbf{M}^m \ . \nonumber \\
\end{eqnarray}
We see that  the requirement of the preservation of the constraint $\mH_T\approx 0$ implies 
an additional constraint $\mH^{II}\equiv 
\frac{2}{\sqrt{h}}\partial_m[\frac{1}{\sqrt{h}}]\mathbf{M}^m+
\frac{1}{h}\partial_m\mathbf{M}^m\approx 0$. 
However then for consistency of theory we should again demand that the constraint 
$\mH^{II}$ is preserved under time evolution with possible new constraint generated. In this way we should continue further and again check requirement of preservation of this constraint and so on.  Of course, there is a possibility to consider extended form of Hamiltonian with all constrains included. In this case the tertiary constraint $
\mH_T$ should be add to the Hamiltonian multiplied by Lagrange multiplier $\lambda^{II}$. Then the consistency of theory  would show that $\mH_T$ is second class constraints with $\mH_T$. Similar situation occurs in case of non-projectable Ho\v{r}ava Lifshitz gravity with space dependent lapse \cite{
	Blas:2009yd,Li:2009bg,Henneaux:2009zb}. Note that this result is possible in principle due to the fact that $\mH_T(\bx)$ in fact  corresponds to $\infty^{n-1}$ constraints. However then the standard counting degrees of freedom would imply that Hamiltonian constraint $\mH_T\approx $ eliminates only one degree of freedom which is inconsistent. In case of non-projectable Ho\v{r}ava-Lifshitz gravity this problem was solved by extension of original gravity with specific term so that the theory possesses two second class constraints making the right counting of physical degrees of freedom. However we mean that such a procedure cannot be applied in case of TDiff invariant gravity and hence it seems that TDiff invariant gravity does not have well defined canonical structure. 

In conclusion we would like to stress that in the special case when $F=1$ the theory has well defined Hamiltonian structure as was shown in our previous paper
\cite{Kluson:2023bkg}. Further, it is possible that the theory has again well defined structure when the matrix $\mM^{ijkl}$ as was defined in our paper is non-invertible. An example such a theory is Weyl transverse gravity \cite{Oda:2016psn}. The canonical analysis of this  theory is currently under study. 

{\bf Acknowledgement:}

This work  is supported by the grant “Dualitites and higher order derivatives” (GA23-06498S) from the Czech Science Foundation (GACR).

\end{document}